\documentstyle[aps,floats,prb]{revtex}

\input epsf

\begin{document}


\twocolumn[\hsize\textwidth\columnwidth\hsize\csname @twocolumnfalse\endcsname

\title{Applications of a tight-binding total
energy method for transition and noble metals:  Elastic
Constants, Vacancies, and Surfaces of Monatomic Metals}

\author{Michael J. Mehl and Dimitrios A. Papaconstantopoulos}
\address{Complex Systems Theory Branch, Naval Research
Laboratory, Washington, D.C. 20375-5345}

\date{\today}
\maketitle

\begin{abstract}
A recent tight-binding scheme provides a method for extending the
results of first principles calculations to regimes involving $10^2
- 10^3$ atoms in a unit cell.  The method uses an analytic set of
two-center, non-orthogonal tight-binding parameters, on-site terms
which change with the local environment, and no pair potential.  The
free parameters in this method are chosen to simultaneously fit band
structures and total energies from a set of first-principles
calculations for monatomic fcc and bcc crystals.  To check the
accuracy of this method we evaluate structural energy differences,
elastic constants, vacancy formation energies, and surface energies,
comparing to first-principles calculations and experiment.  In most
cases there is good agreement between this theory and experiment.
We present a detailed account of the method, a complete set of
tight-binding parameters, and results for twenty-nine of the
alkaline earth, transition and noble metals.
\end{abstract}

\pacs{71.15.-m,62.20.Dc,61.72.Ji,68.35.Bs}
%
]

\section{Introduction}
\label{sec:intro}

Recently, a total energy tight-binding (TB) method was
introduced\cite{cohen94} wherein the parameters were fit to the
total energies and band structures obtained from a limited set of
first-principles calculations.  The method produces very good
structural energy differences, elastic constants, phonon
frequencies, vacancy formation energies, and surface energies for
the transition and noble metals.  In this paper we thoroughly
examine the method for these elements, and for the alkaline earth
metals.

Section~\ref{sec:method} describes the method, including the
analytic expressions for the tight-binding parameters and the scheme
for fitting them to first-principles theory.  The remainder of the
paper discusses the predictions made by this method.
Sec.~\ref{sec:stab} shows that the method correctly predicts the
ordering of several metallic phases, including the $29$ atom
$\alpha$Mn phase, and correctly predicts the energy differences
between the lower phases.  Sec.~\ref{sec:cij} discusses the
calculation of elastic constants.  Since these calculations can also
be performed by first-principles
techniques,\cite{mehl90,mehl91a,mehl93,mehl94} we compare our TB
results to those obtained from first-principles to assess the
accuracy of the method.  We can also use the method to study systems
which have relatively large (20-200 atom) unit cells, a regime which
requires laborious first-principles calculations, but which can be
done very quickly by our TB method.  In Sec.~\ref{sec:vac} we
discuss the structure and energetics of vacancies, and in
Sec.~\ref{sec:surf} we consider surface energetics.  Finally, in
Sec.~\ref{sec:disc} we summarize our results, discuss possible
methods to improve our agreement with first-principles theory or
experiment, and consider the course of future improvements to the
method.

\section{The Tight-Binding Method}
\label{sec:method}

Density functional theory (DFT),\cite{hohen64} using the Kohn-Sham
{\em ansatz} for the kinetic energy,\cite{kohn65} tells us that the
total energy of a system of electrons moving in a solid can be
written in the form
\begin{equation}
E[n ({\bf r})] = \sum_i f(\mu - \varepsilon_i) \varepsilon_i +
F[n({\bf r})] ~ ,
\label{equ:ksdef}
\end{equation}
where $n ({\bf r})$ is the electronic density, $\varepsilon_i$ is
the Kohn-Sham eigenvalue of the $i^{th}$ electronic state, $\mu$ is
the chemical potential, and the sum is over all electronic states of
the system.  Since we will be primarily interested in studying
metals, we take the function $f (\mu-\varepsilon)$ to have the Fermi
function form:\cite{gillan89}
\begin{equation}
f(z) = \frac{1}{1 + e^{\beta z}} ~ ,
\label{equ:fermi}
\end{equation}
where $\beta = 1/(kT)$.  Typically we take $T$ between 2 and 5 mRy.
The functional $F[n ({\bf r})]$ contains the remaining part of the
DFT total energy: the ion-ion interaction energy, the parts of the
Hartree and Exchange-Correlation energy not included in the
eigenvalue sums, and corrections for double counting in the
eigenvalue sums.

Many tight-binding methods use (\ref{equ:ksdef}) to provide a
natural separation between the parts of $E[n ({\bf r})]$ which can
be treated by tight-binding and the parts which must be treated by
other means.
\cite{sigalas94,xu92,goodwin89,carlsson91,sutton88,mercer93,menon94,lathio95}
In these cases the eigenvalue sum is calculated by tight-binding
methods, and $F[n ({\bf r})]$ is approximated by a sum of pair
potentials.

We have developed an alternative method of applying tight-binding to
(\ref{equ:ksdef}),\cite{cohen94} based on the fact that the
Kohn-Sham method allows an arbitrary shift in the potential.  If
this shift is defined to be
\begin{equation}
V_0 = F[n({\bf r})]/N_e ~ ,
\label{equ:v0def}
\end{equation}
where
\begin{equation}
N_e = \sum_i f (\mu - \varepsilon_i)
\end{equation}
is the number of electrons in the system, then the eigenvalues
$\varepsilon_i$ are each shifted by an amount $V_0$, to the new
values
\begin{equation}
\varepsilon'_i = \varepsilon_i + V_0 ~ .
\label{equ:primedef}
\end{equation}
The total energy (\ref{equ:ksdef}) then becomes
\begin{equation}
E[n ({\bf r})] = \sum_i f (\mu'-\varepsilon'_i) \varepsilon'_i ~ ,
\label{equ:shifte}
\end{equation}
where $\mu' = \mu + V_0$ is the shifted chemical potential.

By the density functional theorem,\cite{hohen64} the shifted
eigenvalues $\varepsilon'_i$ can be considered to be functions of
the crystal structure, including volume, primitive lattice vectors,
and internal parameters.  A tight-binding method which reproduces
the $\varepsilon'_i$ over a range of structures will then solve the
total energy problem (\ref{equ:ksdef}) or (\ref{equ:shifte}) without
resort to an additional term.

The two-center Slater-Koster formulation\cite{slater54} of
tight-binding with a non-orthogonal basis breaks the problem into
the calculation of three types of parameters: on-site parameters,
which represent the energy required to place an electron in a
specific orbital, Hamiltonian parameters, which represent the matrix
elements for electrons hopping from one site to another, and overlap
parameters, which describe the mixing between the non-orthogonal
orbitals on neighbor sites.  The eigenvalues $\varepsilon'_i$ can be
determined once the parameters are evaluated for a given structure.
The basic method we use is to give the Slater-Koster parameters
simple algebraic forms, with parameters chosen to reproduce
first-principles results over a wide range of structures.

In an orthogonal tight-binding calculation, applying the shift
(\ref{equ:primedef}) to each eigenvalue would be equivalent to
shifting each diagonal element of the Hamiltonian matrix by an
amount $V_0$.  In the non-orthogonal case the effect is slightly
more complicated.  It is clear, however, that the Hamiltonian and
overlap parameters should not directly depend on the shift $V_0$.
Thus the effect of (\ref{equ:primedef}) can only be accounted for by
changing the on-site parameters.  These parameters must now be
sensitive to the local environment around each atom.  We describe
this local environment by introducing a ``density'' associated with
each atom, defined by
\begin{equation}
\rho_i = \sum_{j \ne i} \exp[-                                   
\lambda_{\tilde{\jmath}\tilde{\imath}}^2 R_{ij}] F_c(R_{ij}) ~ ,
\label{equ:rhodef}                                               
\end{equation}                                                   
where $\tilde{\imath}$ ($\tilde{\jmath}$) denotes the type of atom
on site $i$ ($j$), $\lambda$ is a parameter which will depend on the
atom types, $R_{ij}$ is the distance between atoms $i$ and $j$, and
$F_c(R)$ is a smooth cutoff function,
\begin{equation}
F_c(R) = \{1 + \exp[(R-R_0)/\ell]\}^{-1} ~ ,
\label{equ:cutoff}
\end{equation}
which we use to limit the range of the parameters.  In the
calculations shown here we typically take $R_0 = 14.0$ Bohr and
$\ell = 0.5$ Bohr, which effectively zeros all interactions for
neighbors more than 16.5 Bohr apart.  Depending on the structure and
lattice constant, this radius will include from 80 to 300
neighboring atoms.

Although in principle the on-site terms should have off-diagonal
elements due to the overlap of the on-site wave functions with
neighboring atomic potentials,\cite{mercer94} we follow traditional
practice and only include the diagonal terms, keeping only the terms
corresponding to $s$, $p$, and $d$ orbitals.  The on-site term for
atom $i$ is
\begin{equation}
h_{i \alpha} = a_{\tilde\imath \alpha} + b_{\tilde\imath \alpha}
\rho_i^{2/3} + c_{\tilde\imath \alpha} \rho_i^{4/3} +
d_{\tilde\imath \alpha} \rho^2 ~ ,
\label{equ:onsite}
\end{equation}
where $\alpha = s, p,$ or $d$, and $\rho_i$ is given by
(\ref{equ:rhodef}).  To show the correspondence between this theory
and previous theories, note that $\rho$ in (\ref{equ:rhodef}) has a
form similar to a pair potential.  Thus if (\ref{equ:onsite}) were
linear in $\rho$, we restricted ourselves to an orthogonal
Hamiltonian, and kept $h_{i \alpha}$ independent of the angular
momentum, then (\ref{equ:onsite}) would just be a method of
parametrizing $F[n({\bf r})]$ by a simple pair potential.  Seen in
this light, the density dependent parts of on-site term
(\ref{equ:onsite}) can be regarded as a generalized pair
potential.

In general the Hamiltonian and overlap Slater-Koster parameters may
depend upon the structure and local environment of the atoms, even
in the case of the two-center approximation.  This approach, while
valid, will lead to many complications when we extend the method to
include other crystal structures and binary alloys.  We therefore
constrain the form of these parameters so that they depend only on
the distance between the two atoms.  Both the Hamiltonian and
overlap parameters are assumed to have the same functional form,
\begin{equation}
P_\gamma (R) = (e_\gamma + f_\gamma R) \exp[-g_\gamma^2 R] F_c(R) ~ ,
\label{equ:hopform}
\end{equation}
where $\gamma$ indicates the type of interaction (e.g. $ss\sigma$,
$pd\pi$, etc.), $R$ is the distance between the atoms, and $F_c (R)$
is given by (\ref{equ:cutoff}).

\begin{figure}[hbt]
\epsfysize=2.25in \epsfbox{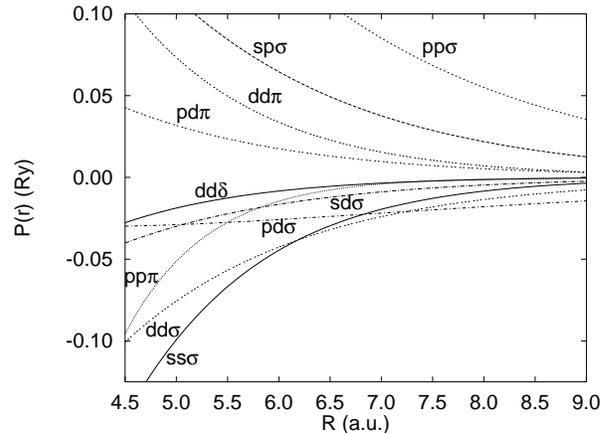}
\caption{Values of the two center Slater-Koster integrals for the
Hamiltonian matrix as a function of distance for molybdenum,
obtained from the tight-binding parameters
(\protect{\ref{equ:hopform}}).  The integrals are labeled according
to the standard notation.\protect{\cite{slater54,papa86}}}
\label{fig:tchop}
\end{figure}

\begin{figure*}[hbt]
\epsfysize=3.0in \epsfbox{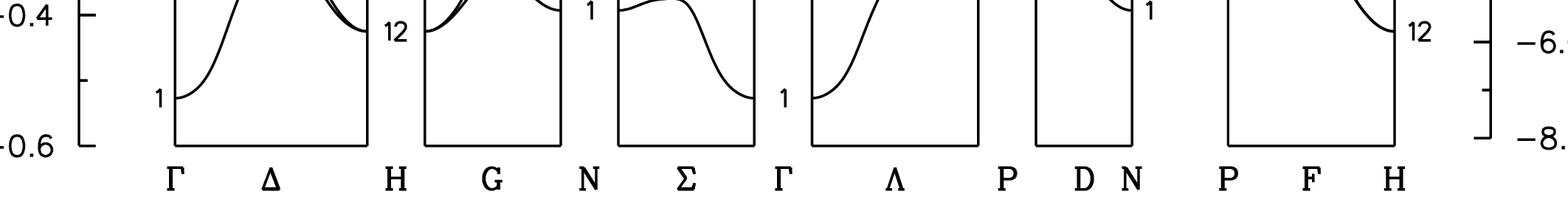}
\caption{The tight-binding electronic band structure of bcc Mo at
the LDA equilibrium lattice constant (3.12\AA).  The energies of the
eigenstates have been uniformly shifted to set the Fermi level to
zero.}
\label{fig:tbbands}
\end{figure*}

In the monatomic case, restricting ourselves to $s$, $p$, and $d$
orbitals, there are three on-site Slater-Koster parameters
(\ref{equ:onsite}), with thirteen parameters required to completely
specify them.  Similarly, there are ten Hamiltonian and ten overlap
parameters (\ref{equ:hopform}), each requiring three parameters.
Thus to completely specify the Slater-Koster tight-binding scheme in
this model requires the determination of at most seventy-three
parameters.  In practice, we only used the $d_{\tilde\imath \alpha}$
found in (\ref{equ:onsite}) in the parametrization of Ti, Zr, and
Hf, so for most elements we had only seventy independent parameters.
The parameters are determined by requiring that the eigenvalues
determined by the Slater-Koster parameters\cite{papa86}
(\ref{equ:rhodef}-\ref{equ:hopform}) reproduce the electronic band
structure (including the eigenvalue shift (\ref{equ:v0def}))
produced by a set of first-principles density functional
calculations, and that the corresponding energies (\ref{equ:shifte})
reproduce the first-principles energies.  For the metals discussed
here we use either the muffin-tin potential Augmented Plane Wave
(APW)\cite{sigalas92} or the full-potential Linearized Augmented
Plane Wave (LAPW)\cite{andersen75,wei85} method to calculate the
electronic band structure and total energies for 4-6 volumes in each
of the fcc and bcc structures, using uniform $k$-point meshes which
include the origin and contain 89 and 55 points, respectively.  In
all cases we used the Hedin-Lundqvist parametrization\cite{hedin71}
of the Local Density Approximation\cite{kohn65} (LDA) to DFT.
Possible spin polarization is ignored.  For a typical transition
metal there are about 4000 eigenvalues and energies in the database.
We use an IMSL package, based on a finite difference
Levenberg-Marquardt algorithm,\cite{more77} to adjust the seventy
(or seventy-three) parameters to reproduce this database by means of
a non-linear least squares fit, with the total energies typically
weighted about 200 times larger than the eigenvalues in a single
band.  We place additional conditions on the minimization as
follows: Since the overlap matrix defined by the parameters
(\ref{equ:hopform}) must, physically, be positive definite, a
function which penalizes small or negative determinant overlap
matrices is included.  While this does not guarantee that our
overlap matrix will remain positive definite during the minimization
process, it does tend to keep the determinant from changing sign,
meaning that an overlap matrix which is originally positive definite
tends to stay that way.  We also restrict the behavior of the
functions (\ref{equ:hopform}).  Physically, we expect both the
Hamiltonian and Overlap parameters to decay steadily and not change
sign for distances $R$ near or greater than the nearest neighbor
distance.  We thus add another contribution to our minimization
function, which penalizes parameter sets ($e_\gamma$, $f_\gamma$,
$g_\gamma$) which cause the either $P_\gamma (r)$ or $P'_\gamma (r)$
to vanish over the range of distances where we are likely to require
evaluation of the Slater-Koster matrix elements.  Finally, we select
our starting parameters guided by those found in previous
work,\cite{papa86} and ensure that the symmetry of the eigenstates
is taken into account.

\begin{figure*}[hbt]
\epsfysize=3.0in \epsfbox{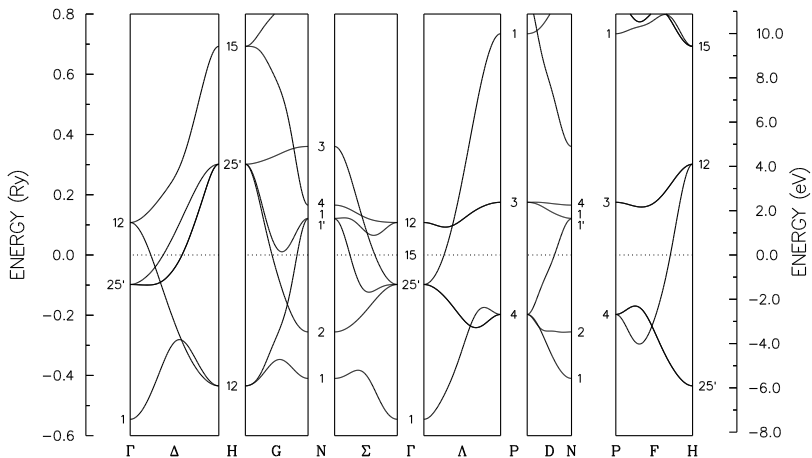}
\caption{The self-consistent first-principles LDA electronic band
structure of bcc Mo at the LDA equilibrium lattice constant
(3.12\AA), as determined by the muffin-tin APW
program.\protect{\cite{sigalas92}} The energies of the eigenstates
have been uniformly shifted to set the Fermi level to zero.}
\label{fig:apwbands}
\end{figure*}

\begin{figure}[hbt]
\epsfysize=2.25in \epsfbox{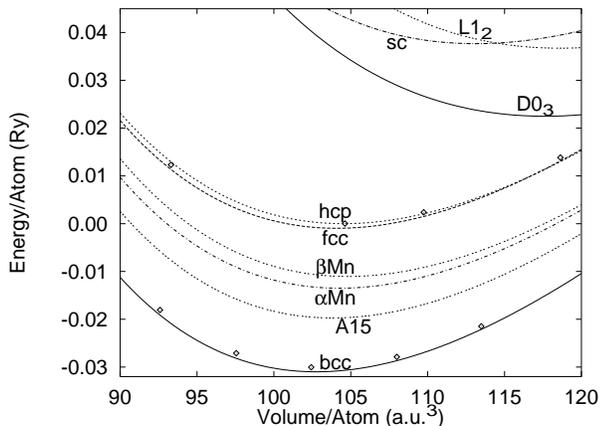}
\caption{The energy/volume behavior of molybdenum in several phases,
using the tight-binding method.  The energy of the diamond structure
is not shown, as its minimum energy is 0.116~Ry above the origin.
The small symbols ($\Diamond$) represent the APW energies of the bcc
and fcc phases.  The construction of the $L1_2$ and $D0_3$ lattices
is discussed in Table~\protect{\ref{tab:lateng}}. The origin of the
energy was chosen arbitrarily.}
\label{fig:moeos}
\end{figure}

As an illustration of the above technique, we present several
results for molybdenum.  Using the APW program, we calculated the
total energy and band structure of Mo at four volumes in the fcc
structure and five volumes in the bcc structure.  The parameters in
(\ref{equ:rhodef}), (\ref{equ:onsite}), and (\ref{equ:hopform}) were
then adjusted to best fit the total energies and the band structure
in the lower six bands of Mo over all nine volumes.  The resulting
two center Slater-Koster parameters (\ref{equ:hopform}) are shown in
Figure~\ref{fig:tchop} for the Hamiltonian matrix elements.  We note
that there is a smooth and monotonic decay of the TB parameters
towards zero.  To assess the quality of the fit, we calculate the
tight-binding band structure of Mo at the LDA equilibrium volume
(3.12\AA) of the bcc structure (Figure~\ref{fig:tbbands}) and
compare it to the first-principles band-structure
(Figure~\ref{fig:apwbands}).\cite{papa86} We see that the band
structures are in close agreement over the entire Brillouin zone and
a wide spectrum of energies, with some deterioration far above the
Fermi level.  We also calculate the energy/volume behavior for a
large number of phases, and compare the fcc and bcc energies to
those obtained from APW calculations, as shown in
Figure~\ref{fig:moeos}.  We see that there is excellent agreement
between the tight-binding and first-principles curves for the fcc
and bcc structures, and that all other structures, including hcp,
are above the bcc ground state.

The above procedure works well for all of the metals described here
except vanadium, where the elastic constant $C_{44}$ is predicted to
be 172~GPa, compared to 30~GPa found experimentally\cite{simmons71}
and 43~GPa when found using the first-principles LAPW method.  This
discrepancy is probably due to the high electronic density of states
in vanadium near the Fermi level.\cite{papa86} To better describe
the electronic structure of vanadium we have made two changes to the
above formalism.  First, we split the $d$ on-site energy
(\ref{equ:onsite}) into contributions from the $t2g$ and $eg$
levels.\cite{papa86} Second, we increase our parameter space by
modifying the Slater-Koster parameter's functional form
(\ref{equ:hopform}) to include a quadratic term,
\begin{equation}
P_\gamma (R) = (e_\gamma + f_\gamma R + \overline{f}_\gamma R^2)
\exp[-g_\gamma^2 R] F_c(R) ~ .
\label{equ:hopformp}
\end{equation}
This procedure adds an additional twenty-four parameters to our
basis set, for a total of ninety-seven.  We reduce this by setting
$d_\alpha = 0$ in (\ref{equ:onsite}) for $\alpha = s$, $p$, $t2g$,
and $eg$.  When we fit the remaining ninety-three parameters, as
above, we reduce the predicted elastic constant $C_{44}$ to 92~GPa
without significantly changing our other predictions.  While this is
not in perfect agreement with experiment, it is a substantial
improvement, so we will use these parameters for vanadium in
preference to the seventy-three parameter set which gives a worse
elastic constant.

We have used this method to determine tight-binding parameters for
most of the alkaline earth, transition and noble metals.  In the
case of the anti-ferromagnetic (Cr and Mn) and magnetic (Fe, Co, and
Ni) elements we restricted our calculations to the non-magnetic
case.  This causes some difficulty when comparing our results to
experiment, as shown below.  The parameters for La are not presented
because its $f$ bands are located in the middle of the $d$
bands\cite{papa86} and therefore it is impossible to fit within our
formalism.  The Appendix lists various methods of obtaining the
parameters used in this paper.

\begin{table*}[htb]
\caption{Equilibrium lattice constants and bulk moduli for the
experimentally observed ground state structures of the elements,
comparing the results of our tight-binding parametrization (TB),
first-principles full potential LAPW results (LAPW), where
available, and experiment (Exp.).\protect{\cite{kittel,donohue}}
Where not referenced, the LAPW results were done for this paper,
using previously developed methods.\protect{\cite{mehl93}}}
\begin{tabular}{lc|ccc|ccc|rrr}
Element & Structure & \multicolumn{3}{c|}{$a$ (\AA)} &
\multicolumn{3}{c|}{$c$ (\AA)} & \multicolumn{3}{c}{$B_0$ (GPa)} \\
\multicolumn{2}{c|}{~} & TB & LAPW & Exp. & TB & LAPW & Exp. & TB &
LAPW & Exp. \\
\tableline
Ca & $fcc$ & \dec 5.30 & \dec 5.34\tablenotemark[1] & \dec 5.58 & $a$ & $a$
& $a$  & 21 & 19\tablenotemark[1] & 15 \\
Sc & $hcp$ & \dec 3.26 & & \dec 3.31 & \dec 4.91 & & \dec 5.27 &
63 & & 44 \\
Ti & $hcp$ & \dec 2.97 & \dec 2.87 & \dec 2.95 & \dec 4.80 & \dec
4.55 & \dec 4.68 & 122 & 120 & 105 \\
V & $bcc$ & \dec 2.94 & \dec 2.93 & \dec 3.03 & $a$ & $a$ & $a$ &
211 & 196 & 162 \\
Cr & $bcc$ & \dec 2.80\tablenotemark[2] & & \dec 2.88 & $a$ & & $a$ &
283 & & 190 \\
Mn & $\alpha$Mn & \dec 8.41\tablenotemark[2] & & \dec 8.91 & $a$ & &
$a$ & 320 & & 60 \\
Fe & $bcc$ & \dec 2.71\tablenotemark[2] & & \dec 2.87 & $a$ & & $a$ &
281 & & 168 \\
Co & $hcp$ & \dec 2.40\tablenotemark[2] & & \dec 2.51 & \dec
3.90\tablenotemark[2] & & \dec 4.07 & 384 & & 191 \\
Ni & $fcc$ & \dec 3.43\tablenotemark[2] & \dec 3.42\tablenotemark[3]
& \dec 3.52 & $a$ & $a$ & $a$ & 268 & 261 & 186 \\
Cu & $fcc$ & \dec 3.52 & \dec 3.52 & \dec 3.61 & $a$ & $a$ & $a$ &
189 & 190 & 137 \\
\tableline
Sr & $fcc$ & \dec 5.73 & & \dec 6.08 & $a$ & $a$ & $a$ & 15 & & 12
\\
Y & $hcp$ & \dec 3.59 & & \dec 3.65 & \dec 5.35 & & \dec 5.73 & 46 &
& 37 \\
Zr & $hcp$ & \dec 2.99 & & \dec 3.23 & \dec 5.57 & & \dec 5.15 & 108
& & 83 \\
Nb & $bcc$ & \dec 3.25 & \dec 3.25 & \dec 3.30 & $a$ & $a$ & $a$ &
187 & 193 & 170 \\
Mo & $bcc$ & \dec 3.12 & \dec 3.12\tablenotemark[3] & \dec 3.15 &
$a$ & $a$ & $a$ & 283 & 291\tablenotemark[3] & 272 \\
Tc & $hcp$ & \dec 2.72 & & \dec 2.74 & \dec 4.34 & & \dec 4.40 & 304
& & 297 \\
Ru & $hcp$ & \dec 2.68 & & \dec 2.71 & \dec 4.26 & & \dec 4.28 & 360
& & 321 \\
Rh & $fcc$ & \dec 3.77 & \dec 3.76 & \dec 3.80 & $a$ & $a$ & $a$ &
306 & 309 & 270 \\
Pd & $fcc$ & \dec 3.85 & \dec 3.85 & \dec 3.89 & $a$ & $a$ & $a$ &
212 & 220 & 181 \\
Ag & $fcc$ & \dec 4.01 & \dec 4.01 & \dec 4.09 & $a$ & $a$ & $a$ &
142 & 142 & 101 \\
\tableline
Ba & $bcc$ & \dec 4.82 & & \dec 5.02 & $a$ & $a$ & $a$ & 10 & & 10
\\
Hf & $hcp$ & \dec 3.07 & & \dec 3.19 & \dec 5.08 & & \dec 5.05 & 111
& & 109 \\
Ta & $bcc$ & \dec 3.30 & \dec 3.24 & \dec 3.30 & $a$ & $a$ & $a$ &
185 & 224 & 200 \\
W & $bcc$ & \dec 3.14 & \dec 3.14 & \dec 3.16 & $a$ & $a$ & $a$ &
319 & 333 & 323 \\
Re & $hcp$ & \dec 2.78 & & \dec 2.76 & \dec 4.39 & & \dec 4.46 & 371
& & 372 \\
Os & $hcp$ & \dec 2.75 & & \dec 2.74 & \dec 4.31 & & \dec 4.32 & 441
& & 418 \\
Ir & $fcc$ & \dec 3.86 & \dec 3.82\tablenotemark[3] & \dec 3.84 & $a$
& $a$ & $a$ & 389 & 401\tablenotemark[3] & 355 \\
Pt & $fcc$ & \dec 3.90 & \dec 3.90 & \dec 3.92 & $a$ & $a$ & $a$ &
318 & 305 & 278 \\
Au & $fcc$ & \dec 4.06 & \dec 4.06 & \dec 4.08 & $a$ & $a$ & $a$ &
187 & 205 & 173 \\
\end{tabular}
\tablenotetext[1]{First-principles LAPW calculations.\cite{mehl91a}}
\tablenotetext[2]{Non-magnetic calculation.  The experimental
structure exhibits some form of magnetism.}
\tablenotetext[3]{First-principles LAPW calculations.\cite{mehl94}}
\label{tab:eos}
\end{table*}

\section{Ground State Behavior and Phase Stability}
\label{sec:stab}

We first test our parameter sets by looking at the equation of state
for the observed ground state of each element.  For the fcc and bcc
non-magnetic cubic crystals, our parameters should reproduce the
first-principles results for these phases because of the fitting
procedure.  For the hcp elements and manganese, these calculations
serve as a tight-binding prediction of the equation of state.  The
case of manganese has been discussed in more detail
elsewhere.\cite{mehl95}

Table \ref{tab:eos} shows the equilibrium lattice constants and bulk
moduli for all of the metals studied here, comparing them to
first-principles LAPW calculations\cite{mehl91a,mehl94} and to
experiment.\cite{kittel,donohue} For the non-magnetic cubic
crystals, the equilibrium lattice constant is within 1\% of the
first-principles LDA value for all of the cubic elements except
tantalum and iridium, and in these cases the discrepancy is less
than 2\%.  Similarly, the equilibrium bulk moduli are in good
agreement with the first-principles results, within 10\% for all
non-magnetic cubic elements except tantalum, where the error is
17\%.

The hexagonal close packed phases were not fit to first-principles
results.  As a result, the equations of state predicted by the
tight-binding model are not as accurate as for the cubic lattices.
The largest error is for zirconium, where $a$ is 7\% smaller than
experiment and $c$ is 8\% larger.  Yttrium and hafnium also show
large discrepancies between the tight-binding model and experiment,
and the $c$ for titanium is 2.5\% larger than experiment.  The
lattice constants for the other elements are within 2\% of
experiment, consistent with the errors we would find in
first-principles calculations.  We conclude that the tight-binding
method is nearly as good as first-principles LDA methods in
determining the structural parameters of hcp metals.

\begin{table*}[htb]
\caption{Relative energies per atom of several structures for each
of the metals examined by the tight-binding model discussed in the
text.  The energy of the experimental ground state structure is
arbitrarily set to zero.  All energies are calculated at the
equilibrium volume found by the tight-binding fit, and are expressed
in mRy.  Below the common name of each phase is its {\em
Strukturbericht} designation.}
\begin{tabular}{lcccccccccc}
Element & \multicolumn{10}{c}{Structure} \\
\tableline
 & fcc & bcc & hcp & diamond & sc & $\alpha$Mn & $\beta$Mn &
$\beta$W & AuCu$_3$ & AlFe$_3$ \\
 & A1 & A2 & A3 & A4 & A$_h$ & A12 & A13 & A15 &
L1$_2$\tablenotemark[1] & D0$_3$\tablenotemark[2] \\
\tableline
Ca & \dec 0.0 & \dec 2.8 & \dec 0.5 & \dec 83.5 & \dec 13.3 & \dec
2.1 & \dec 3.1 & \dec 0.7 & \dec 19.8 & \dec 24.4 \\
Sc & \dec 4.7 & \dec 8.5 & \dec 0.0 & \dec 101.1 & \dec 34.3 & \dec
10.2 & \dec 11.0 & \dec 19.4 & \dec 39.7 & \dec 33.9 \\
Ti & \dec 5.2 & \dec 9.1 & \dec 0.0 & \dec 157.7 & \dec 64.9 & \dec
16.7 & \dec 15.4 & \dec 22.4 & \dec 56.1 & \dec 63.8  \\
V & \dec 19.6 & \dec 0.0 & \dec 19.6 & \dec 180.8 & \dec 76.0 &
\dec 11.2 & \dec 12.7 & \dec 12.2 & \dec 62.1 & \dec 53.4 \\
Cr & \dec 28.6 & \dec 0.0 & \dec 30.6 & \dec 49.1 & \dec 118.6 &
\dec 20.3 & \dec 19.7 & \dec 16.2 & \dec 94.3 & \dec 87.2 \\
Mn & \dec 7.5 & \dec 14.3 & \dec 2.9 & \dec 44.2 & \dec 90.3
& \dec 0.0 & \dec 1.5 & \dec 7.0 & \dec 87.6 & \dec 66.2 \\
Fe & \dec -27.9 & \dec 0.0 & \dec -35.0 & \dec 79.8 & \dec 54.5 &
\dec -23.1 & \dec -25.0 & \dec -8.1 & \dec 4.5 & \dec 20.4 \\
Co & \dec -2.9 & \dec 21.0 & \dec 0.0 & \dec 135.0 & \dec 8.2 & & &
\dec 19.2 & \dec 72.7 & \dec 75.2 \\
Ni & \dec 0.0 & \dec 8.0 & \dec 2.8 & \dec 142.3 & \dec 75.9 &
\dec 7.7 & \dec 4.6 & \dec 15.0 & \dec 48.7 & \dec 57.1 \\
Cu & \dec 0.0 & \dec 3.5 & \dec 1.2 & \dec 70.8 & \dec 24.7 &
\dec 6.2 & \dec 5.4 & \dec 10.1 & \dec 14.7 & \dec 22.0 \\
\tableline
Sr & \dec 0.0 & \dec  2.1 & \dec 1.0 & \dec  73.0 & \dec 22.0 &
\dec 3.2 & \dec 3.2 & \dec 1.4 & \dec 22.2 & \dec 25.4 \\
Y  & \dec 2.4 & \dec  8.8 & \dec 0.0 & \dec  97.9 & \dec 432.2 &
\dec 8.6 & \dec 10.4 & \dec 17.3 & \dec 31.5 & \dec 27.5 \\
Zr & \dec 1.1 & \dec 2.9 & \dec 0.0 & \dec 112.7 & \dec 123.3 & 
\dec 8.6 & \dec 8.0 & \dec 8.0 & \dec 18.7 & \dec 26.2 \\
Nb & \dec 26.1 & \dec 0.0 & \dec 22.7 & \dec 181.9 & \dec 68.6 &
\dec 17.1 & \dec 21.1 & \dec 15.9 & \dec 69.5 & \dec 59.6 \\
Mo & \dec 30.0 & \dec 0.0 & \dec 31.0 & \dec 147.3 & \dec 68.7 &
\dec 17.5 & \dec 19.9 & \dec 11.3 & \dec 67.7 & \dec 53.5 \\
Tc & \dec 6.1 & \dec 23.3 & \dec 0.0 & \dec 72.9 & \dec 56.3 &
\dec 0.2 & \dec 4.5 & \dec 13.3 & \dec 64.6 & \dec 35.4 \\
Ru & \dec 1.0 & \dec 44.6 & \dec 0.0 & \dec 134.5 & \dec 106.6 &
\dec 16.3 & \dec 14.6 & \dec 34.7 & \dec 103.0 & \dec 82.7 \\
Rh & \dec 0.0 & \dec 31.7 & \dec 5.1 & \dec 159.8 & \dec 96.0 &
\dec 16.3 & \dec 7.4 & \dec 29.5 & \dec 81.2 & \dec 88.3 \\
Pd & \dec 0.0 & \dec 9.9 & \dec 2.5 & \dec 148.6 & \dec 84.9 &
\dec 9.2 & \dec 6.1 & \dec 12.8 & \dec 50.1 & \dec 72.8 \\
Ag & \dec 0.0 & \dec 3.3 & \dec 0.6 & \dec 66.1 & \dec 24.1 &
\dec 6.0 & \dec 5.3 & \dec 8.6 & \dec 15.7 & \dec 25.2 \\
\tableline
Ba & \dec 0.9 & \dec 0.0 & \dec 0.1 & \dec 85.3 & \dec 23.6 &
\dec 2.8 & \dec 2.9 & \dec 2.2 & \dec 27.0 & \dec 30.7 \\
Hf & \dec 1.2 & \dec 7.1 & \dec 0.0 & \dec 380.2 & \dec 113. & \dec
22.9 & \dec 20.6 & \dec 20.8 & \dec 110.2 & \dec 143.3 \\
Ta & \dec 24.7 & \dec 0.0 & \dec 25.3 & \dec 188.6 & \dec 63.2
& \dec 8.9 & 14.0 & \dec 9.8 & \dec 65.6 & \dec 54.2 \\
W & \dec 36.0 & \dec 0.0 & \dec 10.1 & \dec 159.3 & \dec 113.7 &
\dec 18.2 & \dec 25.7 & \dec 12.4 & \dec 134.2 & \dec 94.2 \\
Re & \dec 13.4 & \dec 27.7 & \dec 0.0 & \dec 28.6 & \dec 55.6 &
\dec 0.2 & \dec 1.8 & \dec 19.1 & \dec 61.0 & \dec 22.1 \\
Os & \dec 8.1 & \dec 64.3 & \dec 0.0 & \dec 11.2 & \dec 58.6 &
\dec 17.8 & \dec 17.4 & \dec 42.4 & \dec 66.2 & \dec 33.6 \\
Ir & \dec 0.0 & \dec 50.1 & \dec 8.5 & \dec 83.9 & \dec 83.5 &
\dec 23.8 & \dec 19.9 & \dec 41.8 & \dec 64.0 & \dec 75.6 \\
Pt & \dec 0.0 & \dec 9.7 & \dec 4.4 & \dec 168.7 & \dec 78.3 &
\dec 14.4 & \dec 12.8 & \dec 25.9 & \dec 48.5 & \dec 78.5 \\
Au & \dec 0.0 & \dec 1.4 & \dec 0.6 & \dec 96.1 & \dec 24.1 &
\dec 8.7 & \dec 8.4 & \dec 15.2 & \dec 17.5 & \dec 32.3 \\
\end{tabular}
\tablenotetext[1]{Calculations using the L1$_2$ (AuCu$_3$) lattice,
placing the metal atoms on the copper sites and leaving vacancies on
the gold sites.}
\tablenotetext[2]{Calculations using the D0$_3$ (AlFe$_3$) lattice,
placing the metal atoms on the iron sites and leaving vacancies on
the aluminum sites.}
\label{tab:lateng}
\end{table*}

\begin{figure}[htb]
\epsfysize=2.25in \epsfbox{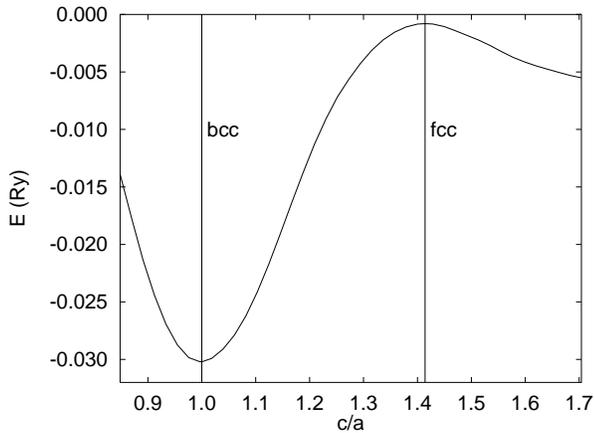}
\caption{The tight-binding calculation of the energy of molybdenum,
at the experimental equilibrium volume, under a tetragonal strain
(the Bain path\protect{\cite{bain24}}), as a function of $c/a$.  The
vertical lines denote the positions of the bcc and fcc lattices.}
\label{fig:bain}
\end{figure}

To be successful, the tight-binding method must also show that the
experimental ground state structure is, in fact, the ground state
predicted by the method.  We have calculated energy versus volume
curves for several common crystal structures.  As an example,
Figure~\ref{fig:moeos} shows the resulting energy/volume curves for
molybdenum.  The results for all of the elements are summarized in
Table~\ref{tab:lateng}, which shows the equilibrium energy of each
of these phases expressed as the difference in energy between that
phase and the equilibrium energy of the experimental ground state.
The $c/a$ ratio of the hcp phase has been chosen to minimize the
tight-binding energy at each volume to obtain the hcp energies.  The
tight-binding method correctly predicts the ground state structure
for all metals except the ferromagnetic metals iron and cobalt.
This is understandable, since we have not included
spin-polarization.  Especially satisfying is the fact the method
finds the correct ground states of the hcp metals and
manganese,\cite{mehl95} since these phases are not included in the
fit of the tight-binding parameters.

\begin{figure}[hbt]
\epsfysize=2.25in \epsfbox{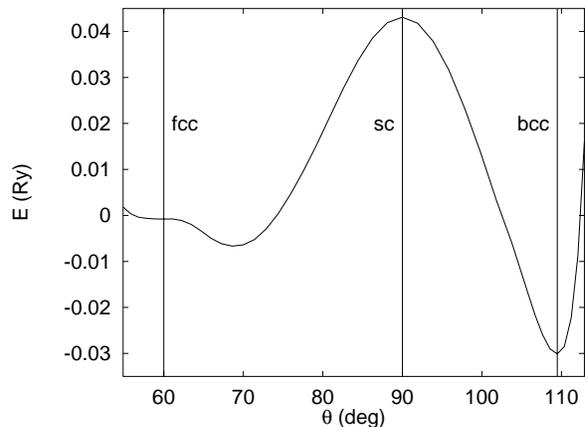}
\caption{The tight-binding calculation of the energy of molybdenum,
at the experimental equilibrium volume, under a trigonal strain of
the lattice given by (\protect{\ref{equ:triglat}}), as a function of
the angle between the primitive vectors, $\theta$.  The positions of
the bcc, simple cubic, and fcc lattices are denoted by the vertical
lines.}
\label{fig:trig}
\end{figure}

We can also assess the stability of the tight-binding parameters by
studying crystal structures of lower symmetry.  Thus
Figure~\ref{fig:bain} shows the energy of molybdenum under a volume
conserving tetragonal strain, the so-called Bain path,\cite{bain24}
at the experimentally observed equilibrium volume of the bcc phase.
The energy is properly a minimum for the bcc lattice, where $c/a =
1$, and attains a local maximum at the fcc structure ($c/a =
\sqrt2$), indicating that $C_{11}-C_{12} < 0$, so that the fcc
structure is unstable.  Similarly, Figure~\ref{fig:trig} shows the
energy for a monatomic lattice undergoing a trigonal strain with the
primitive vectors
\begin{equation}
\left( \begin{array}{c} {\bf a}_1 \\ {\bf a}_2 \\ {\bf a}_3
\end{array} \right) = \left( \begin{array}{ccc} \alpha & \beta &
\beta \\ \beta & \alpha & \beta \\ \beta & \beta & \alpha
\end{array} \right) \left(
\begin{array}{c} \hat{x} \\ \hat{y} \\ \hat{z} \end{array} \right) ~
,
\label{equ:triglat}
\end{equation}
where $\alpha$ and $\beta$ are chosen so that the volume is the
experimentally observed volume of the bcc lattice, and the angle
between the primitive vectors is given by $\theta$, where
\begin{equation}
\cos \theta = \frac{(2 \alpha + \beta) \beta}{\alpha^2 + 2 \beta^2}
~ .
\end{equation}
The vectors (\ref{equ:triglat}) represent an fcc lattice when
$\theta = 60\deg$, a simple cubic lattice when $\theta = 90\deg$,
and a bcc lattice when $\theta = 109.47\deg$.  The tight-binding
calculations in Figure~\ref{fig:trig} show that the bcc phase is,
indeed, the minimum energy structure.  Also, the simple cubic phase
has a local maximum in the energy, indicating that it is elastically
unstable with $C_{44} < 0$.

\begin{table}[hbt]
\caption{Elastic constants for cubic elements.  All elements for
which we have constructed a set of tight-binding parameters and
which have a cubic ground state (except manganese) are presented
here.  Comparison is made between the results of our tight-binding
parametrization (TB), first-principles full potential LAPW results
(LAPW), where available, and experiment
(Exp.).\protect{\cite{simmons71}} Where unreferenced, the LAPW
results were done for this paper, using previously developed
methods.\protect{\cite{mehl93}} Calculations were performed at the
experimental volume.}
\begin{tabular}{l|rrr|rrr|rrr}
 & \multicolumn{3}{c|}{$C_{11}$} & \multicolumn{3}{c|}{$C_{12}$} &
\multicolumn{3}{c}{$C_{44}$} \\
 & TB & LAPW & Exp & TB & LAPW & Exp & TB & LAPW & Exp \\
\tableline
Ca & 15 & 17\tablenotemark[1] & 16\tablenotemark[2] & 10 &
10\tablenotemark[1] & 12\tablenotemark[2] & 4 & 14\tablenotemark[1]
& 8\tablenotemark[2] \\
V  &  224 & 205 & 228 &  106 & 111 & 119 &   92 &  30 &  43 \\
Cr &  432 &     & 346 &   88 &     &  66 &  250 &     & 100 \\
Fe &   -4 &     &     &  227 &     &     &  180 &     &     \\
Ni &  256 &     &     &  142 &     &     &   86 &     &     \\
Cu &  161 & 156 & 168 &  108 & 106 & 121 &   55 &  80 &  75 \\
Sr &    8 &     &  15 &    3 &     &   6 &   -3 &     &  10 \\
Nb &  204 & 230 & 246 &  137 & 122 & 139 &   34 &  25 &  29 \\
Mo & 453 & 468\tablenotemark[1] & 450\tablenotemark[3] & 147 &
149\tablenotemark[1] & 173\tablenotemark[3] & 120 &
98\tablenotemark[1] & 125\tablenotemark[3] \\
Rh &  491 & 433 & 413 &  171 & 185 & 194 &  260 & 206 & 184 \\
Pd &  233 & 218 & 227 &  163 & 172 & 176 &   63 &  74 &  72 \\
Ag &  133 & 122 & 124 &   86 &  90 &  93 &   42 &  52 &  46 \\
Ba &    9 &     &     &    7 &     &     &   13 &     &     \\
Ta &  275 & 256 & 261 &  140 & 154 & 157 &   78 &  67 &  82 \\
W  &  529 & 527 & 523 &  170 & 194 & 203 &  198 & 147 & 160 \\
Ir & 694 & 621\tablenotemark[1] & 590 & 260 & 256\tablenotemark[1] &
249 & 348 & 260\tablenotemark[1] & 262 \\
Pt &  380 & 381 & 347 &  257 & 189 & 251 &   71 &  83 &  76 \\
Au &  184 & 200 & 189 &  154 & 173 & 159 &   43 &  33 &  42 \\
\end{tabular}
\tablenotetext[1]{First-principles LAPW calculations.\cite{mehl94}}
\tablenotetext[2]{Experiment\cite{smithells83}}
\tablenotetext[3]{Experiment\cite{feather63}}
\label{tab:cubcij}
\end{table}

\section{Elastic Constants}
\label{sec:cij}

Elastic constants are calculated by imposing an external strain on
the crystal, relaxing any internal parameters to obtain the energy
as a function of the strain, and numerically solving for the elastic
constants as the curvature of the energy versus strain
curve.\cite{mehl90,mehl91a,mehl93,mehl94} We have calculated the
elastic constants for the ground state phases of all of the elements
in this study except manganese.  This procedure serves two purposes:
it tests the mechanical stability of the tight-binding ground state;
and it assesses the ability of the method to determine properties
not included in the fit of the tight-binding parameters.

The elastic constants of the cubic materials (except manganese) are
shown in Table~\ref{tab:cubcij}.  This table also shows the elastic
constants obtained from first-principles LAPW
computations\cite{mehl94} and
experiment.\cite{simmons71,smithells83,feather63} All calculations
were performed at the experimental volume.  The accuracy of the
elastic constants $C_{11}$ and $C_{12}$ is remarkably good.  For the
twelve elements for which we have first-principles results, none of
the tight-binding elastic constants differs from the
first-principles results by more than 15\%.  For comparison, the
first-principles results also differ from the experimental results
by less than 15\%, a measure of the accuracy of the LDA in
determining these elastic constants.  Thus the tight-binding elastic
constants are not significantly worse than those computed from
first-principles.  Considering only the non-magnetic materials, the
root-mean-square (RMS) deviation of the tight-binding $C_{11}$
compared to experiment is 38~GPa, while the RMS deviation of
$C_{12}$ is 15~GPa.

The errors in $C_{44}$ are somewhat larger, the RMS deviation from
experiment being 49~GPa.  In this case, if we exclude calcium and
strontium, discussed below, the maximum deviation of
first-principles results from experiment is 22\%, in molybdenum.
The maximum deviation of the tight-binding elastic constant from the
first-principles result is in vanadium, where the error is 114\%,
even when we use the quadratic Slater-Koster parametrization
(\ref{equ:hopformp}) and split the on-site $d$ terms.  As noted
above, this is probably due to the high value of the density of
states of vanadium near the Fermi level.\cite{papa86} Excluding
vanadium, the maximum deviation is 36\%, in niobium.  For
comparison, the first-principles results differ from experiment by
at most 22\%, in molybdenum.  Hence, except for the anomaly of
vanadium, the tight binding results are only slightly worse than the
first-principles results in the determination of elastic constants.

Calculation of $C_{44}$ in the alkaline earth metals is difficult
because the lattice is extremely soft.  The error in calcium is
71\%, compared to first-principles results, and 50\% compared to
experiment.  For strontium the situation is even worse, since we
predict the fcc lattice to be unstable to any strain which is
related to $C_{44}$.  In this regard, however, it is important to
note that even many of the best LDA elastic constant calculations in
Table~\ref{tab:cubcij} have an absolute error of more than 10 GPa.
Since the experimental values of $C_{44}$ for both calcium and
strontium are less than 10 GPa, it is understandable that we cannot
reproduce the experimental elastic constants for the alkaline earth
metals with any degree of accuracy.

We also note that this method correctly reproduces the sign of the
elastic constant $C_{12} - C_{44}$, even in rhodium and iridium,
where it is negative, though the predicted magnitude $|C_{12} -
C_{44}|$ is much larger than observed experimentally and in
first-principles calculations for these two materials.  The negative
sign cannot be obtained from the standard Embedded Atom Method
(EAM),\cite{voter94} though it can be found in the Modified Embedded
Atom Method (MEAM).\cite{baskes92}

\begin{table}[hbt]
\caption{Elastic constants for hexagonal close-packed elements.  All
elements for which we have constructed a set of tight-binding
parameters and which have a non-magnetic $hcp$ ground state are
presented here.  Comparison is made between the results of our
tight-binding parametrization (TB) and experiment
(Exp).\protect{\cite{smithells83}} The tight-binding results include
internal relaxation.  Calculations were performed at the
experimental volume, but at the $c/a$ ratio which minimized the
energy for that volume.  Note that in a hexagonal crystal, $C_{66} =
(C_{11} - C_{12})/2$.}
\begin{tabular}{l|rr|rr|rr|rr|rr}
 & \multicolumn{2}{c|}{$C_{11}$} & \multicolumn{2}{c|}{$C_{12}$} &
\multicolumn{2}{c|}{$C_{13}$} & \multicolumn{2}{c|}{$C_{33}$} &
\multicolumn{2}{c}{$C_{44}$} \\
 & TB & Exp & TB & Exp & TB & Exp & TB & Exp & TB & Exp \\
\tableline
Sc & 73  &  99 &  30 &  40 &  31 &  29 &  78 & 107 &  25 & 28 \\
Ti & 171 & 160 &  58 &  90 &  46 &  60 & 203 & 181 &  64 & 47 \\
Y  & 45  &  78 &  15 &  29 &  13 &  20 &  48 &  77 &  17 & 24 \\
Zr & 102 & 144 &  67 &  74 &  65 &  67 &  99 & 166 &   4 & 33 \\
Tc & 477 &     & 196 &     & 172 &     & 505 &     & 127 & \\
Ru & 642 & 563 & 170 & 188 & 129 & 168 & 706 & 624 & 214 & 181 \\
Re & 559 & 616 & 283 & 273 & 250 & 206 & 621 & 683 & 136 & 161 \\
Os & 754 &     & 272 &     & 274 &     & 831 &     & 232 & \\
\end{tabular}
\label{tab:hexcij}
\end{table}

We have also calculated elastic constants for all of the
non-magnetic transition metals which take the hcp structure.
Hexagonal close packed crystals have two atoms per unit cell, and
the internal parameter describing the atomic positions is free to
move when we strain the crystal to calculate elastic constants.
Using the tight-binding approach, it is relatively simple to
determine the value of the internal parameter which minimizes the
total energy for a given external strain, and then find the related
elastic constant by standard techniques.  This is more difficult for
first principles calculations, so we have not calculated LAPW
elastic constants for hexagonal materials.  The tight-binding
elastic constants are compared to experiment\cite{smithells83} in
Table~\ref{tab:hexcij}.  For simplicity in interpreting the results,
we calculated the elastic constants using the tight-binding $c/a$
ratio at the experimental volume.  The hcp tight-binding elastic
constants show larger relative deviations from experiment than found
in the cubic crystals.  The largest deviation, 88\%, occurs in the
calculation of $C_{44}$ in zirconium, where we predict a value of
4~GPa, compared to the 33~GPa found by experiment.  The RMS
deviations of the tight-binding elastic constants from experiment
are 46~GPa for $C_{11}$, 17~GPa for $C_{12}$, 25~GPa for $C_{13}$,
54~GPa for $C_{33}$, and 22~GPa for $C_{44}$.  The RMS deviation
over all of the hcp elastic constants is 36~GPa.  When compared to
the RMS deviation for cubic materials (37~GPa), we see that on
average the hcp elastic constants are as accurate as those
calculated in the cubic case.

\begin{table}[hbt]
\caption{Tight-binding vacancy formation energies compared to
first-principles calculations and experiment.  Energies were
computed using a 108 atom supercell.  Calculations with the atoms at
the primitive lattice sites in the crystal (Fixed) and allowing
relaxation around the vacancy (Relaxed) are shown.  First-principles
calculations of the vacancy formation energy are given in the column
labeled ``LDA''.  The experimental column shows a range of energies
if several experiments have been tabulated.  Otherwise the estimated
error in the experiment is given.}
\begin{tabular}{lcccc}
 & \multicolumn{4}{c}{Vacancy Formation Energy} \\
Element & \multicolumn{2}{c}{Tight-Binding} & LDA &
Experiment\tablenotemark[1] \\
 & Fixed & Relaxed & &  \\
\tableline
Cu & 1.29 & 1.18 & 1.41\tablenotemark[2] 1.29\tablenotemark[3] &
1.28 - 1.42 \\
Nb & 2.84 & 2.82 &      & 2.65 $\pm$ 0.3 \\
Mo & 2.63 & 2.46 &      & 3.0 - 3.6 \\
Rh & 3.39 & 3.35 & 2.26\tablenotemark[3] & 1.71\tablenotemark[4] \\
Pd & 2.46 & 2.45 & 1.57\tablenotemark[2]     & 1.85 $\pm$ 0.25 \\
Ag & 1.31 & 1.24 & 1.20\tablenotemark[2] 1.06\tablenotemark[3] &
1.11 - 1.31 \\
Ta & 3.17 & 2.95 &      & 2.9 $\pm$ 0.4 \\
W  & 6.86 & 6.43 &      & 4.6 $\pm$ 0.8 \\
Ir & 2.19 & 2.17 &      & 1.97\tablenotemark[4] \\
Pt & 2.79 & 2.79 &      & 1.35 $\pm$ 0.09 \\
Au & 1.24 & 1.12 &      & 0.89 $\pm$ 0.04
\end{tabular}
\tablenotetext[1]{Experimental energies as tabulated by
Schaefer\cite{schaefer87} unless otherwise noted.}
\tablenotetext[2]{First-principles calculations of Dederichs {\em
et al.}.\cite{dederichs91}}
\tablenotetext[3]{First-principles calculations of Polatoglou {\em
et al.}\cite{polato93}}
\label{tab:vac}
\tablenotetext[4]{Experimental energy tabulated by de Boer {\em et
al.}\cite{deboer88}}
\end{table}

\section{Vacancies}
\label{sec:vac}

Vacancy formation energies are most easily determined by the
supercell total-energy method.\cite{mehl91} One atom in the
supercell is removed, and neighboring atoms are allowed to relax
around this vacancy while preserving the symmetry of the lattice.
The great advantage of the tight-binding method over
first-principles calculations is that we can do the calculation in a
very large supercell, eliminating the possibility of vacancy-vacancy
interactions.  We find that a supercell containing 108 atoms is
sufficient to eliminate the interaction.  The vacancy formation
energy is given by
\begin{equation}
E_{vac} (V) = E_{sc} (N-1,1;V) - (N-1) E_{bulk} (V/N) ~ ,
\end{equation}
where $E_{sc} (M,Q;V)$ is the total energy of a supercell of volume
V containing $M$ atoms and $Q$ vacancies, and $E_{bulk} (V)$ is the
energy per atom of the bulk metal at a volume $V$ per atom.  We use
the experimental lattice constant to set the volume of the system,
since under experimental conditions the lattice constant of a metal
containing isolated vacancies will be the lattice constant of the
bulk metal.

We have computed the vacancy formation energy of several of the
cubic transition and noble metals using this tight binding method,
including relaxation around the vacancy.  The results are presented
in Table~\ref{tab:vac}, where we also compare to first principles
results\cite{dederichs91,polato93} and
experiment.\cite{schaefer87,deboer88} The calculated vacancy
formation energies of niobium, silver, tantalum, and iridium are in
excellent agreement with experiment.  The tight-binding
parametrizations of rhodium, tungsten, and platinum give the poorest
vacancy formation energies, 1--2~eV above the experimental value.

We note that the $L1_2$ and $D0_3$ lattices described in
Table~\ref{tab:lateng} can be considered vacancy-containing
supercells with $N=4$.  Though the vacancy-vacancy interaction in
these cells is quite large, fitting the TB parameters
(\ref{equ:rhodef}-\ref{equ:hopform}) to first-principles results for
these lattices will improve the vacancy formation energy
calculation.  We have yet to implement this procedure.

\begin{table*}[p]
\caption{Surface energies, calculated from the tight-binding theory
(TB), by first principles Local Density methods (LDA), by the
embedded atom method (EAM), or by the modified embedded atom method
(MEAM), compared to experiment.  Energies are given in units of
$J/m^2$.}
\begin{tabular}{lcccccc}
Element & Orientation & TB & LDA & EAM &
MEAM\tablenotemark[1] & Experiment \\
\tableline
   & (111) & 3.34 &  &  & 1.81 & 2.6\tablenotemark[2] \\
V  & (110) & 2.13 &  & 1.68\tablenotemark[3] & 1.71 & \\
   & (100) & 3.04 &  & 1.83\tablenotemark[3] & 2.49 & \\
\tableline
   & (111) & 1.73 & 1.94\tablenotemark[4] & 1.17\tablenotemark[5] & 1.41 &
1.77\tablenotemark[6]  \\
Cu & (110) & 2.04 &  & 1.40\tablenotemark[5] & 1.64 & \\
   & (100) & 1.93 &  & 1.28\tablenotemark[5] & 1.65 & \\
\tableline
   & (111) & 2.44 &  &  & 2.02 & 2.3\tablenotemark[2] \\
Nb & (110) & 1.54 & 2.36\tablenotemark[7] ~ 2.9\tablenotemark[8] &
 1.81\tablenotemark[3] & 1.87 & \\
   & (100) & 2.37 & 2.86\tablenotemark[7] ~ 3.1\tablenotemark[8] &
 1.97\tablenotemark[3] & 2.02 & \\
\tableline
   & (111) & 2.84 &  &  & 1.86 & 2.9\tablenotemark[2] \\
Mo & (110) & 3.04 & 3.14\tablenotemark[7] & 2.13\tablenotemark[3]
 & 1.93 & \\
   & (100) & 2.12 & 3.52\tablenotemark[7] & 2.28\tablenotemark[3]
 & 2.12 & \\
\tableline
   & (111) & 2.46 & 2.54\tablenotemark[4] ~ 2.53\tablenotemark[7] &
& 2.60 & 2.6\tablenotemark[2] \\
Rh & (110) & 2.71 & 2.88\tablenotemark[7] & & 2.92 & \\
   & (100) & 2.57 & 2.81\tablenotemark[7] & & 2.90 & \\
\tableline
   & (111) & 1.57 & 1.64\tablenotemark[7] & 1.22\tablenotemark[5] &
1.38 & 2.00\tablenotemark[2] \\
Pd & (110) & 1.86 & 1.97\tablenotemark[7] ~ 2.5\tablenotemark[8] ~
2.70\tablenotemark[9] & 1.49\tablenotemark[5] & 1.67 & \\
   & (100) & 1.75 & 1.86\tablenotemark[7] ~ 2.3\tablenotemark[8] ~
1.04\tablenotemark[9] & 1.37\tablenotemark[5] & 1.66 & \\
\tableline
   & (111) & 1.14 &  1.21\tablenotemark[7] ~ 1.21\tablenotemark[4] &
0.62\tablenotemark[5] & 1.09 & 1.32\tablenotemark[6] \\
Ag & (110) & 1.42 & 1.26\tablenotemark[7] ~ 1.4\tablenotemark[8]~ &
0.77\tablenotemark[5]& 1.27 & \\
   & (100) & 1.29 & 1.21\tablenotemark[7] ~ 1.3\tablenotemark[8]~ &
0.71\tablenotemark[5] & 1.22 & \\
\tableline
   & (111) & 3.14 & &   & 2.31 & 2.78\tablenotemark[2] \\
Ta & (110) & 2.05 & & 1.80\tablenotemark[3]  & 2.17 & \\
   & (100) & 3.00 & & 1.99\tablenotemark[3]  & 3.29 & \\
\tableline
  & (111) & 6.75 & &      & 2.25 & 2.99\tablenotemark[6] \\
W & (110) & 4.30 & & 2.60\tablenotemark[3] & 2.23 & \\
  & (100) & 6.7 & 5.54\tablenotemark[10] & 2.81\tablenotemark[3] &
2.65 & 6\tablenotemark[11] \\
\tableline
   & (111) & 2.59 & &      & 2.84 & 3.0\tablenotemark[2] \\
Ir & (110) & 3.19 & &      & 3.06 & \\
   & (100) & 2.95 & &      & 2.91 & \\
\tableline
   & (111) & 2.51 & & 1.44\tablenotemark[5] & 1.66 & 2.49\tablenotemark[2] \\
Pt & (110) & 2.97 & & 1.75\tablenotemark[5] & 2.13 & \\
   & (100) & 2.83 & & 1.65\tablenotemark[5] & 2.17 & \\
\tableline
   & (111) & 1.48 & & 0.79\tablenotemark[5] & 0.89 & 1.54\tablenotemark[6] \\
Au & (110) & 1.85 & & 0.98\tablenotemark[5] & 1.12 & \\
   & (100) & 1.69 & & 0.92\tablenotemark[5] & 1.08 & \\
\end{tabular}
\tablenotetext[1]{Modified Embedded-Atom calculations.\cite{baskes92}}
\tablenotetext[2]{Experimental determination of the surface energy
of an ``average'' face using an approximate extrapolation to
T~=~0.\cite{baskes92}}
\tablenotetext[3]{Embedded-Atom calculations for bcc
metals.\cite{guellil92}}
\tablenotetext[4]{Full-Potential Linearized Muffin-Tin Orbital
(Full Potential-LMTO) calculation, using seven layer
slabs.\cite{polato93}}
\tablenotetext[5]{Embedded-Atom calculations for fcc
metals.\cite{foiles86}}
\tablenotetext[6]{Experimental determination of the surface energy
of an ``average'' face at T~=~0.\cite{tyson77}}
\tablenotetext[7]{FP-LMTO calculation, using seven layer
slabs.\cite{methfessel92}}
\tablenotetext[8]{Full-Potential LAPW calculation, using nine
layer slabs.\cite{weinert89}}
\tablenotetext[9]{Local orbital pseudo-potential calculation, using
a three layer slab.\cite{tomanek91}}
\tablenotetext[10]{Full-Potential Linearized Augmented Plane Wave
(LAPW) calculation for the ideal surface, using a five layer
slab.\cite{singh86}}
\tablenotetext[11]{Experimental result for the (100)
surface.\cite{cordwell69}}
\label{tab:surf}
\end{table*}

\section{Surfaces}
\label{sec:surf}

The tight-binding model can be used to calculate surface energies by
the supercell technique.  A slab of metal is formed by cleaving the
crystal along the desired plane, creating two identical free
surfaces.  The distance between the two surfaces in increased,
creating a set of slabs which repeat periodically in the direction
perpendicular to the surfaces.  Of course, we must take care to
separate the slabs by a large region of vacuum so that the electrons
on one slab cannot hop to a neighboring slab.  The slabs must be
thick enough so that the atoms at the center of the slab have the
properties (e.g., local electronic density of states) of atoms in
the bulk material, and so that the two surfaces on the same slab
cannot interact with each other.  For low index ((100), (110), and
(111)) faces of the non-magnetic fcc and bcc metals we find that a
slab containing 25 atomic layers is sufficient to meet these
criteria.  We also converge the calculations with respect to the
$k$-point mesh.  Depending on the surface and underlying bulk
structure this takes between 19 and 91 $k$-points in the two
dimensional Brillouin zone.  We estimate numerical error in the TB
surface energies to be about 0.1 J/m$^2$.

The surface energy, expressed as the energy required to create a
unit area of new surface, is then given by the formula
\begin{equation}
E_{surf} = \frac1{2A} ( E_{slab} - N E_{bulk} )
\end{equation}
where A is the area occupied by one unit cell on the surface of the
slab, $E_{slab}$ is the total energy of the slab, $N$ is the number
of atoms in the unit cell, and $E_{bulk}$ is the energy of one atom
in the bulk at the lattice constant of the atoms in the interior of
the slab.  For simplicity, the calculations were done using the bulk
equilibrium lattice parameters, with no relaxation or reconstruction
at the surface.  Our calculations indicate that relaxation {\em
energies} are on the order of 1-10\% of the total surface energy, so
neglecting relaxation will not significantly alter our conclusions.
Of course, relaxation and reconstruction are necessary in order to
understand surface properties.  We will address these issues in a
future paper.

We have calculated the formation energies for bulk terminated (i.e.,
unrelaxed and unreconstructed) low index ((100), (110), and (111))
surfaces of the non-magnetic fcc and bcc transition and noble
metals.  Table~\ref{tab:surf} compares these results to
experiment,\cite{tyson77,cordwell69} first-principles
calculations,\cite{polato93,tomanek91,methfessel92,weinert89,singh86}
and atomistic models.\cite{baskes92,foiles86,guellil92} Note that
except for the (100) surface of tungsten, the experimental
results\cite{baskes92,tyson77} are for an ``average'' surface, and
some are extrapolated to T~=~0 from high
temperatures.\cite{baskes92}

The results for the fcc metals are particularly gratifying.  In all
cases we find $E_{(111)} < E_{(100)} < E_{(110)}$.  Thus
close-packed surfaces are the most stable for the fcc metals.  The
TB surface energies are uniformly larger, and closer to experiment,
than those obtained by the embedded atom method
(EAM),\cite{foiles86} which is known to underestimate surface
energies in fcc metals.  Our surface energies are generally closer
to experiment than those obtained by the Modified Embedded Atom
Method (MEAM),\cite{baskes92} and in good agreement with
first-principles calculations of the surface energy of
copper(111),\cite{polato93} and rhodium.\cite{methfessel92} Our
calculations are also in the range of the surface energies predicted
by several first-principles calculations for palladium and
silver.\cite{tomanek91,methfessel92,weinert89}

Results for bcc metals are also good.  For all of the unrelaxed
surfaces we find $E_{(110)} < E_{(100)} < E_{(111)}$.  The energies
for niobium and molybdenum are somewhat low compared to both
first-principles calculations\cite{methfessel92,weinert89} and
experiment, but the only serious discrepancy is the average surface
energy of tungsten, which is smaller than our values by a factor of
one-third.  At the moment this is not particularly worrisome,
because experimental estimates of the surface energy of tungsten
range from 1.68~J/m$^2$ to 4.50~J/m$^2$.\cite{tyson75} The latter
number is, in fact, in range of our calculation for the (110)
surface.  In addition, the results for the energy of the (100)
surface of tungsten are in good agreement with both first-principles
results\cite{singh86} and experiment.\cite{cordwell69}.  In most
cases the tight-binding surface energies are greater than either of
the corresponding EAM or MEAM values.  Where this is not the case
the energy difference is slight.

\section{Summary}
\label{sec:disc}

We have extended the tight-binding method\cite{cohen94} to include
most of the alkaline earth, transition, and noble metals.
Tight-binding parameters are determined by doing a least squares fit
to simultaneously reproduce the total energy and electronic
structure determined by first-principles at several volumes of the
fcc and bcc lattices.  In all cases except the ferromagnetic metals
the method correctly predicts the ground state structure, even when
the ground state is hcp or $\alpha$Mn.

The method was tested against first-principles results and
experiment by the calculation of elastic constants, vacancy
formation energies, and surface energies.  Elastic constants for
non-magnetic cubic materials are, in general, in good agreement with
both LAPW calculations and experiment, though we have some problems
with $C_{44}$ in vanadium.  Since the tight-binding method does not
require a specific sign for $C_{12}-C_{44}$ we get reasonable
elastic constants even for rhodium and iridium.  Elastic constants
for the hcp phase are nearly as accurate as those calculated in the
cubic phase.

Compared to experiment, the accuracy of the vacancy formation energy
predictions range from good to poor, with niobium, silver, tantalum,
and iridium all in excellent agreement with experiment, and rhodium,
tungsten, and platinum much larger than experiment.

Surface energies for both fcc and bcc crystals are in good agreement
with the available first-principles and experimental data.  Energies
for fcc crystals are uniformly larger than the corresponding EAM
energies, in agreement with experiment.

Improvements to the tight-binding method are under investigation.
In particular, we are examining other forms for the parametrization
of the on-site terms (\ref{equ:onsite}) and the Slater-Koster terms
(\ref{equ:hopform}).  We are also investigating the effect of
two-center Slater-Koster integrals on the on-site terms, as
suggested by Mercer and Chou.\cite{mercer94} Since these terms
properly account for the crystal-field splitting of the on-site
terms, they may be useful in obtaining even more accurate results.
Finally, we have begun the process of extending the tight-binding
method to binary systems.\cite{papacon95}

\acknowledgments We thank Warren Pickett, David Singh and Mihalis
Sigalas for many helpful discussions.  This work is partially
supported by the United States Office of Naval Research.

\appendix

\section*{The Tight-Binding Parameters}

The tight-binding parameters used in this paper are available from
the AIP Physics Auxiliary Publication Service,
{\tt http://www.aip.org/epaps/epaps.html},
as noted in a footnote to this paper, on the World Wide Web at {\tt
http://cst-www.nrl.navy.mil/bind}, or via email at {\tt
mehl@dave.nrl.navy.mil}.


\begin{table*}[p]
\caption{Tight-binding parameters
(\protect{\ref{equ:rhodef}}-\protect{\ref{equ:hopform}}) for
Calcium, Scandium, Titanium, Vanadium, and Chromium.  The equation
numbers refer to the parametrization forms discussed in the text.
Note that the $t2g$ and $eg$ on-site parameters are identical for
all elements except vanadium.}
\begin{tabular}{lrrrrr}
Parameter & \multicolumn{1}{c}{Calcium} & \multicolumn{1}{c}{Scandium}
& \multicolumn{1}{c}{Titanium} & \multicolumn{1}{c}{Vanadium} & \multicolumn{1}{c}{Chromium} \\
\tableline
 & \multicolumn{5}{c}{On-Site Parameters Eqs. (\ref{equ:rhodef}) and
(\ref{equ:onsite})} \\
\tableline
$\lambda$ &  1.12775E$+$00 & 1.16238E$+$00 & 1.30248E$+$00 & 1.578632$+$00 & 1.59665E$+$00             \\
$a_s$     &  2.43912E$-$01 & 3.91870E$-$01 & $-$1.53116E$-$02 & 1.24744E$-$01 & 3.57154E$-$01          \\
$b_s$     &  $-$1.73150E$+$01 & $-$1.29635E$+$01 & 1.82222E$+$01 & $-$3.46195E$+$01 & $-$8.83822E$+$00  \\
$c_s$     &  5.96806E$+$02 & 4.06795E$+$02 & 2.12586E$+$03 & 6.62479E$+$04 & 3.91628E$+$04             \\
$d_s$     &  0.00000E$+$00 & 0.00000E$+$00 & 3.35194E$+$03 & 0.00000E$+$00 & 0.00000E$+$00             \\
$a_p$     &  1.67761E$-$01 & 1.08349E$-$01 & 3.73246E$-$01 & 5.09467E$-$01 & 6.65273E$-$01             \\
\end{tabular}
\label{tab:param1}
\end{table*}

\begin{table*}
\begin{tabular}{lrrrrr}
$e_{dd\sigma}$ & 4.76025E$+$00 & 9.13632E$-$01 & $-$8.83254E$-$01 & $-$1.38932E$+$00 & $-$1.44185E$+$00        \\
$f_{dd\sigma}$ & $-$9.11415E$-$01 & $-$2.84308E$-$01 & $-$1.06276E$-$02 & 4.40043E$-$02 & 8.41726E$-$02        \\
$\overline{f}_{dd\sigma}$ & 0.00000E$+$00 & 0.00000E$+$00 & 0.00000E$+$00 & $-$9.20172E$-$03 & 0.00000E$+$00     \\
$g_{dd\sigma}$ & 8.17314E$-$01 & 7.58386E$-$01 & 7.48868E$-$01 & 8.12780E$-$01 & 7.92347E$-$01                 \\
$e_{dd\pi}$    & $-$8.83124E$-$01 & 8.09225E$-$01 & 2.14981E$+$00 & 2.44958E$+$00 & 3.37805E$+$00              \\
$f_{dd\pi}$    & 5.71572E$-$01 & 3.86043E$-$01 & 4.84974E$-$02 & 1.90258E$-$01 & 3.41379E$-$02                 \\
$\overline{f}_{dd\pi}$ & 0.00000E$+$00 & 0.00000E$+$00 & 0.00000E$+$00 & 1.08500E$+$00 & 0.00000E$+$00     \\
$g_{dd\pi}$    & 8.35313E$-$01 & 8.67158E$-$01 & 8.78521E$-$01 & 9.45587E$-$01 & 9.58150E$-$01                 \\
\end{tabular}
\end{table*}

\begin{table*}[p]
\caption{Tight-binding parameters
(\protect{\ref{equ:rhodef}}-\protect{\ref{equ:hopform}}) for
Manganese, Iron, Cobalt, Nickel, and Copper.
The equation numbers refer to the parametrization forms discussed in
the text.}
\begin{tabular}{lrrrrr}
Parameter & \multicolumn{1}{c}{Manganese} & \multicolumn{1}{c}{Iron}
& \multicolumn{1}{c}{Cobalt} & \multicolumn{1}{c}{Nickel} & \multicolumn{1}{c}{Copper} \\
\tableline
 & \multicolumn{5}{c}{On-Site Parameters Eqs. (\ref{equ:rhodef}) and
(\ref{equ:onsite})} \\
\tableline
$\lambda$ & 1.65076E$+$00 & 1.61592E$+$00 & 1.43580E$+$00 & 1.43847E$+$00 & 1.47292E$+$00              \\
$a_s$     & 2.82713E$-$01 & 3.13939E$-$01 & 3.11786E$-$01 & 2.26020E$-$01 & 2.13398E$-$02              \\
$b_s$     & $-$2.44625E$+$01 & $-$1.50341E$+$02 & 3.03424E$+$01 & 2.85451E$+$01 & 6.93939E$+$01        \\
$c_s$     & 8.90173E$+$04 & 5.60132E$+$04 & 5.76787E$+$02 & $-$1.40381E$+$02 & $-$2.44787E$+$03        \\
$d_s$     & 0.00000E$+$00 & 0.00000E$+$00 & 0.00000E$+$00 & 0.00000E$+$00 & 0.00000E$+$00              \\
$a_p$     & 7.92094E$-$01 & 7.50228E$-$01 & 4.87203E$-$01 & 4.69576E$-$01 & 5.55994E$-$01              \\
\end{tabular}
\label{tab:param2}
\end{table*}

\begin{table*}
\begin{tabular}{lrrrrr}
   & \multicolumn{5}{c}{Overlap Parameters Eq. (\ref{equ:hopform})}  \\
\tableline
$e_{ss\sigma}$ & $-$3.87663E$+$00 & 2.08692E$+$00 & 3.37736E$+$01 & 4.38609E$+$01 & $-$4.62973E$+$00           \\
$f_{ss\sigma}$ & 2.21239E$+$00 & 1.50952E$+$00 & $-$6.02198E$+$00 & $-$1.03415E$+$00 & 9.47334E$+$00           \\
$g_{ss\sigma}$ & 8.66857E$-$01 & 9.38173E$-$01 & 1.14555E$+$00 & 1.17048E$+$00 & 1.14594E$+$00                 \\
$e_{sp\sigma}$ & 4.67005E$+$00 & 6.61325E$+$00 & $-$3.59473E$+$00 & $-$4.81962E$+$00 & $-$7.76506E$+$00        \\
$f_{sp\sigma}$ & $-$1.26965E$+$00 & $-$2.39708E$+$00 & 3.78546E$-$01 & 1.25105E$-$01 & $-$7.02631E$-$01        \\
$g_{sp\sigma}$ & 7.60452E$-$01 & 8.42816E$-$01 & 9.00865E$-$01 & 9.51048E$-$01 & 9.74543E$-$01                 \\
$e_{pp\sigma}$ & 9.65378E$+$00 & 7.79047E$+$00 & $-$2.08360E$-$01 & $-$3.93614E$-$01 & $-$8.67146E$-$01        \\
\end{tabular}
\end{table*}

\begin{table*}[p]
\caption{Tight-binding parameters
(\protect{\ref{equ:rhodef}}-\protect{\ref{equ:hopform}}) for
Strontium, Yttrium, Zirconium, Niobium, and Molybdenum.
The equation numbers refer to the parametrization forms discussed in
the text.}
\begin{tabular}{lrrrrr}
Parameter & \multicolumn{1}{c}{Strontium} & \multicolumn{1}{c}{Yttrium}
& \multicolumn{1}{c}{Zirconium} & \multicolumn{1}{c}{Niobium} & \multicolumn{1}{c}{Molybdenum} \\
\tableline
 & \multicolumn{5}{c}{On-Site Parameters Eqs. (\ref{equ:rhodef}) and
(\ref{equ:onsite})} \\
\tableline
$\lambda$ & 1.14372E$+$00 & 1.27304E$+$00 & 9.92096E$-$01 & 1.34109E$+$00 & 1.35071E$+$00       \\
$a_s$     & 1.77264E$-$01 & 7.28123E$-$02 & 4.56862E$-$01 & 2.50727E$-$02 & 8.66826E$-$02       \\
$b_s$     & $-$1.31630E$+$01 & 9.32506E$+$00 & $-$1.73980E$+$00 & 5.47470E$+$01 & 6.34195E$+$01 \\
$c_s$     & 1.37270E$+$03 & $-$4.92331E$+$02 & $-$2.37835E$+$01 & 5.11141E$+$03 & 9.36509E$+$02 \\
$d_s$     & 0.00000E$+$00 & 0.00000E$+$00 & 2.45157E$+$02 & 0.00000E$+$00 & 0.00000E$+$00       \\
$a_p$     & 1.76186E$-$01 & 4.14728E$-$01 & 3.26016E$-$01 & 5.88564E$-$01 & 7.20981E$-$01       \\
\end{tabular}
\label{tab:param3}
\end{table*}

\begin{table*}
\begin{tabular}{lrrrrr}
   & \multicolumn{5}{c}{Overlap Parameters Eq. (\ref{equ:hopform})}  \\
\tableline
$e_{ss\sigma}$ & 1.32808E$+$01 & $-$1.32457E$+$00 & $-$1.30200E$+$00 & 3.19400E$-$01 & 3.05256E$+$02        \\
$f_{ss\sigma}$ & $-$1.27851E$+$00 & 2.71913E$+$00 & 2.30501E$+$00 & $-$6.04851E$-$04 & $-$3.52015E$+$01     \\
$g_{ss\sigma}$ & 7.07779E$-$01 & 8.08101E$-$01 & 8.99383E$-$01 & 7.44630E$-$01 & 1.26105E$+$00              \\
$e_{sp\sigma}$ & 3.73826E$+$04 & 5.24175E$+$02 & 4.30489E$+$02 & $-$1.98466E$+$03 & 1.32679E$+$03           \\
$f_{sp\sigma}$ & $-$5.69590E$+$03 & $-$1.01379E$+$02 & $-$1.10041E$+$02 & 1.59179E$+$02 & $-$4.00480E$+$02  \\
$g_{sp\sigma}$ & 1.16729E$+$00 & 1.02133E$+$00 & 1.12375E$+$00 & 1.34929E$+$00 & 1.33379E$+$00              \\
$e_{pp\sigma}$ & 6.30614E$+$02 & 2.32538E$+$01 & 8.90408E$-$01 & 4.02407E$+$00 & 1.10983E$+$00              \\
\end{tabular}
\end{table*}

\begin{table*}[p]
\caption{Tight-binding parameters
(\protect{\ref{equ:rhodef}}-\protect{\ref{equ:hopform}}) for
Technetium, Ruthenium, Rhodium, Palladium, and Silver.
The equation numbers refer to the parametrization forms discussed in
the text.}
\begin{tabular}{lrrrrr}
Parameter & \multicolumn{1}{c}{Technetium} & \multicolumn{1}{c}{Ruthenium}
& \multicolumn{1}{c}{Rhodium} & \multicolumn{1}{c}{Palladium} & \multicolumn{1}{c}{Silver} \\
\tableline
 & \multicolumn{5}{c}{On-Site Parameters Eqs. (\ref{equ:rhodef}) and
(\ref{equ:onsite})} \\
\tableline
$\lambda$ & 1.35605E$+$00 & 1.34253E$+$00 & 1.34056E$+$00 & 1.27733E$+$00 & 1.22495E$+$00          \\
$a_s$     & 1.56969E$-$01 & 8.93160E$-$02 & 9.12651E$-$02 & 9.21895E$-$02 & 1.71377E$-$01          \\
$b_s$     & 4.96528E$+$01 & 4.12031E$+$01 & 4.48975E$+$01 & 2.66437E$+$01 & 8.37704E$+$00          \\
$c_s$     & 6.87483E$+$02 & 1.29943E$+$03 & 8.96404E$+$02 & 2.45427E$+$02 & $-$1.77374E$+$01       \\
$d_s$     & 0.00000E$+$00 & 0.00000E$+$00 & 0.00000E$+$00 & 0.00000E$+$00 & 0.00000E$+$00          \\
$a_p$     & 7.27706E$-$01 & 6.70655E$-$01 & 6.20365E$-$01 & 5.05324E$-$01 & 6.05047E$-$01          \\
\end{tabular}
\label{tab:param4}
\end{table*}

\begin{table*}
\begin{tabular}{lrrrrr}
   & \multicolumn{5}{c}{Overlap Parameters Eq. (\ref{equ:hopform})}  \\
\tableline
$e_{ss\sigma}$ & 2.97103E$+$02 & 1.27671E$+$01 & $-$1.80513E$+$03 & 1.17693E$+$04 & 5.19751E$+$00             \\
$f_{ss\sigma}$ & $-$3.76325E$+$01 & 8.94019E$-$02 & 0.00000E$+$00 & $-$5.39387E$+$02 & $-$1.81289E$-$01       \\
$g_{ss\sigma}$ & 1.25946E$+$00 & 1.95994E$+$00 & 2.21069E$+$00 & 1.62329E$+$00 & 8.60426E$-$01                \\
$e_{sp\sigma}$ & 1.44349E$+$03 & $-$1.05981E$+$04 & $-$1.12235E$+$04 & $-$1.53705E$+$04 & $-$2.28701E$+$00    \\
$f_{sp\sigma}$ & $-$3.79690E$+$02 & $-$2.72224E$+$02 & 0.00000E$+$00 & 2.32248E$+$03 & $-$1.61325E$-$01       \\
$g_{sp\sigma}$ & 1.30455E$+$00 & 1.62336E$+$00 & 1.61109E$+$00 & 1.50175E$+$00 & 7.74383E$-$01                \\
$e_{pp\sigma}$ & 7.57425E$-$01 & 1.58799E$+$01 & $-$5.49048E$+$00 & 2.19627E$+$00 & $-$3.78981E$+$00          \\
\end{tabular}
\end{table*}	
		
\begin{table*}[p]	
\caption{Tight-binding parameters
(\protect{\ref{equ:rhodef}}-\protect{\ref{equ:hopform}}) for
Barium, Hafnium, Tantalum, and Tungsten.
The equation numbers refer to the parametrization forms discussed in
the text.}	
\begin{tabular}{lrrrr}
Parameter & \multicolumn{1}{c}{Barium} & \multicolumn{1}{c}{Hafnium}
& \multicolumn{1}{c}{Tantalum} & \multicolumn{1}{c}{Tungsten} \\
\tableline	
 & \multicolumn{4}{c}{On-Site Parameters Eqs. (\ref{equ:rhodef}) and
(\ref{equ:onsite})} \\
\tableline	
$\lambda$ & 1.11634E$+$00 &  1.19630E$+$00 &  1.27675E$+$00 &  1.32674E$+$00       \\
$a_s$     & 1.50138E$-$01 &  9.10131E$-$01 &  3.19292E$-$01 &  4.33397E$-$01       \\	
$b_s$     &$-$1.53745E$+$01 &  5.29294E$+$00 &  3.66642E$-$01 &  1.42840E$+$00     \\	
$c_s$     & 1.56954E$+$03 &  5.04222E$+$02 &  4.42664E$+$02 &  4.00732E$+$02       \\	
$d_s$     & 0.00000E$+$00 &  4.95952E$+$02 &  0.00000E$+$00 &  0.00000E$+$00       \\	
$a_p$     & 1.28894E$-$01 &  4.65965E$-$01 &  3.59226E$-$01 &  4.27055E$-$01       \\
\end{tabular}
\label{tab:param5}
\end{table*}

\begin{table*}
\begin{tabular}{lrrrrr}
   & \multicolumn{4}{c}{Overlap Parameters Eq. (\ref{equ:hopform})}  \\
\tableline
$e_{ss\sigma}$ & 1.83722E$+$01 &  1.18231E$+$01 &  4.95296E$+$00 & $-$1.01186E$+$03      \\
$f_{ss\sigma}$ &$-$1.43245E$+$00 &  2.96234E$+$00 &  0.00000E$+$00 &  2.64629E$+$02      \\
$g_{ss\sigma}$ & 7.19225E$-$01 &  9.46583E$-$01 &  8.52620E$-$01 &  1.25307E$+$00        \\
$e_{sp\sigma}$ & 4.44201E$+$04 &  1.72009E$+$00 & $-$4.73578E$+$02 & $-$3.78579E$-$01    \\
$f_{sp\sigma}$ &$-$6.40797E$+$03 & $-$2.35586E$+$00 &  0.00000E$+$00 & $-$1.09561E$-$01  \\
$g_{sp\sigma}$ & 1.14903E$+$00 &  8.62716E$-$01 &  1.31345E$+$00 &  7.26138E$-$01        \\
$e_{pp\sigma}$ &$-$7.35150E$+$02 &  1.54031E$+$01 & $-$1.11120E$-$01 & $-$7.94937E$+$01  \\
\end{tabular}
\end{table*}
		
\begin{table*}[p]	
\caption{Tight-binding parameters
(\protect{\ref{equ:rhodef}}-\protect{\ref{equ:hopform}}) for
Rhenium, Osmium, Iridium, Platinum, and Gold.
The equation numbers refer to the parametrization forms discussed in
the text.}	
\begin{tabular}{lrrrrr}
Parameter & \multicolumn{1}{c}{Rhenium} & \multicolumn{1}{c}{Osmium}
& \multicolumn{1}{c}{Iridium} & \multicolumn{1}{c}{Platinum} & \multicolumn{1}{c}{Gold} \\
\tableline	
 & \multicolumn{5}{c}{On-Site Parameters Eqs. (\ref{equ:rhodef}) and
(\ref{equ:onsite})} \\
\tableline	
$\lambda$ & 1.41690E+00 & 1.43787E+00 & 1.48862E+00 & 1.48637E+00 & 1.37357E+00 \\
$a_s$     & 1.25774E$-$01 & 2.19235E$-$01 & 2.58676E$-$01 & 5.83156E$-$03 & 9.52842E$-$02 \\
$b_s$     & 4.32410E+01 & 2.32773E+01 & 8.99491E+01 & 2.05116E+02 & 4.03569E+01 \\
$c_s$     &$-$2.65528E+03 &$-$1.49716E+02 &$-$5.38470E+03 &$-$2.04105E+04 &$-$3.05963E+03 \\
$d_s$     & 0.00000E+00 & 0.00000E+00 & 0.00000E+00 & 0.00000E+00 & 0.00000E+00 \\
$a_p$     & 6.08286E$-$01 & 7.15539E$-$01 & 6.49919E$-$01 & 8.17823E$-$01 & 5.75191E$-$01 \\
\end{tabular}
\label{tab:param6}
\end{table*}

\begin{table*}
\begin{tabular}{lrrrrr}
   & \multicolumn{5}{c}{Overlap Parameters Eq. (\ref{equ:hopform})}  \\
\tableline
$e_{ss\sigma}$ &$-$7.65992E+02 & 3.42921E+01 & 8.03977E+00 & 8.48032E+00 & 5.49189E+00 \\
$f_{ss\sigma}$ & 2.20348E+02 &$-$1.86163E+00 &$-$1.31689E+00 &$-$1.21344E+00 &$-$5.81247E$-$01 \\
$g_{ss\sigma}$ & 1.23551E+00 & 1.05666E+00 & 8.31100E$-$01 & 8.73589E$-$01 & 8.01636E$-$01 \\
$e_{sp\sigma}$ & 2.45545E+03 &$-$9.08968E+01 & 1.86281E+03 & 2.07666E+03 & 1.47488E+00 \\
$f_{sp\sigma}$ &$-$5.52012E+02 & 7.58654E+00 &$-$5.45115E+02 &$-$4.87117E+02 &$-$1.70493E+00 \\
$g_{sp\sigma}$ & 1.33722E+00 & 1.12351E+00 & 1.41339E+00 & 1.42828E+00 & 8.83162E$-$01 \\
$e_{pp\sigma}$ &$-$1.87751E+01 & 2.24577E$-$01 &$-$4.81893E+02 &$-$1.59104E+04 & 4.39256E+00 \\
\end{tabular}
\end{table*}

\newpage


\end{document}